\documentstyle[sprocl,epsf,rotate]{article}
\input{epsfig}
\def\Journal#1#2#3#4{{#1} {\bf #2}, #3 (#4)}

\def\PLB{{\em Phys. Lett.}  B}

\def\PRD{{\em Phys. Rev.} D}

\def\be{\begin{equation}}
\def\ee{\end{equation}}
\def\bea{\begin{eqnarray}}
\def\eea{\end{eqnarray}}

\begin{document}
\title{SOME POSSIBILITIES FOR CHARM STUDIES AT B-FACTORIES}
\author{ E. GOLOWICH}
\address{Physics Department, University of Massachusetts,
Amherst,\\ MA 01003, USA}
\maketitle\abstracts{To provide focus to my review 
of charm theory at the {\it B-Physics and CP-Violation} 
conference, I consider theoretical issues related to 
possible charm studies at $B$-factories. A few topics of 
interest to the author are stressed -- rare decays, 
CP-violating asymmetries and ${\bar D}^0 - D^0$ mixing.}
\section{Introduction}\label{sec;intro}
To me, the CKM matrix is first and foremost an object 
of phenomenological study.  When I look at it, it is natural 
to associate with each matrix element at least one 
{\it experiment}, as in 
$$
\left[ \begin{array}{lll} 
V_{\rm ud}&V_{\rm us}&V_{\rm ub}\\ 
V_{\rm cd}&V_{\rm cs}&V_{\rm cb}\\
V_{\rm td}&V_{\rm ts}&V_{\rm tb}
\end{array}\right] \leftrightarrow 
\left[ \begin{array}{lll} 
\beta \ {\rm decay}& K \to \pi \ell {\bar \nu} 
  &B_d \to X_{\rm u} \ell {\bar \nu}\\
\nu N \to \mu {\bar \mu} X & D \to {\bar K} \ell {\bar \nu} 
& B \to D^* \ell {\bar \nu} \\
B_d - {\bar B}_d & b \to s \gamma &  t \to b W  
\end{array}\right] \ \ .
$$
The gradual and painstaking determination 
of the CKM matrix over a long period of time is 
a remarkable accomplishment of our field, and 
hopefully it will attain a level of completion 
from the B-factory studies to come.  

The emphasis of $B$-factories will be on the physics of $b$ quarks and 
deservedly so.  Already, there is discovery of a $\Delta B = 2$ process 
(${\overline B}_d - B_d$ mixing) and an EW-penguin decay 
($B \to K^*~ \gamma$).  The future observation of CP violations (CPV) in $B$
decays would top the list.

However, $B$ factories will provide opportunities for doing $c$ quark 
studies as well.  I would hope that the following items be given 
some consideration in the planning now underway:   
\begin{itemize}
\item Search for flavor-changing neutral current (FCNC) charm 
decays, especially 
\subitem $D\to M \gamma$ ($M$ denotes a $J>0$ non-charm meson), 
\subitem $D \to M' \ell^+\ell^-$ ($M'$ denotes a non-charm meson), 
\item Tighten the limit on CPV asymmetries in $D^\pm$ decay, 
\item Tighten the limit on $D^0-{\bar D}^0$ mixing.  
\end{itemize}
Although many more possible studies of charm systems come to mind, 
this talk will concentrate on just these few.  

\section{Current Status of Weak Decay Studies}\label{sec:wda}
I shall begin by giving a very brief overview of the current 
status of the electroweak sector of charm physics.
An important `truth' evidenced in weak decays of charm 
is the relative importance of the strong interactions. 
This is already signalled by the large number of identifiable 
hadron resonances in the charm region.  For example, in the 
most recent listing of the Particle Data Group, 
I count twelve non-strange and six strange resonances in the 
$1700 \to 2000$~MeV mass region.~\cite{Pd96}   
\begin{table}[t]
\caption{Measured Phases.\label{tab:phase}}
\vspace{0.4cm}
\begin{center}
\begin{tabular}{c|c}Mode & $\delta_I - \delta_{I'}$ \\ \hline\hline
$K {\bar K}$ & ($28^{+9}_{-17})^o$ \\ 
$K \pi$ & $(97^{+12}_{-13})^o$ \\
$K^* \pi$ & $(90 \pm 16)^o$ \\ 
$K \rho$ & $(10 \pm 47)^o$ \\
$K^* \rho$ & $(33 \pm 57)^o$ \\
$\pi \pi $ & $(81 \pm 10)^o$ \\
\hline\hline
\end{tabular}
\end{center}
\end{table}
\subsection{Experimental Studies}\label{subsec:fsi}
If QCD effects are significant in the charm region, 
one expects phases to be prevalent in weak decay 
amplitudes due to effects of final state interactions (FSI). 
It is a sign of the maturity in experimental studies that 
such phases are routinely extracted from a study of isospin 
sum rules.  Table~\ref{tab:phase} summarizes the current state of 
affairs.~\cite{cleo1,teb1} An example is the recent 
CLEO analysis of the $D\to K{\bar K}$ system~\cite{cleo1} in 
which the physical decay amplitudes are expressed in terms of isospin, 
\be
A_{+-} = {A_1 + A_0 \over \sqrt{2}}\ , \quad 
A_{+0} = \sqrt{2}A_1 \ , \quad A_{00} = {A_1 - A_0 \over \sqrt{2}} \ \ .
\ee
This allows extraction of the phase difference 
$\delta_{{\overline K}K} \equiv \delta_{I=1} - \delta_{I=0}$ from 
the measured decay widths, 
\be
\cos\delta_{{\overline K}K} = {\Gamma_{+-} - \Gamma_{00} \over 
\left[ \Gamma_{+0}(2\Gamma_{+-} + 2\Gamma_{00} - \Gamma_{+0}) 
\right]^{1/2} } \ \ .
\ee

\subsection{Theoretical Studies}\label{subsec:thy1}
The list of theory papers on charm decays is very long. 
A recent and comprehensive study of $D$ decays is that of 
Buccella, Lusignoli and Pugliese,~\cite{Bu95}
who study a large number of modes and, most notably, 
include FSI effects in their fits.  As a result, each of 
their weak decay amplitudes has a phase.  This provides them 
with the ability to estimate both CPV asymmetries and 
$\Delta \Gamma_D / \Gamma_D$.  We shall consider each of these in 
due course.
\begin{table}[t]
\caption{DCSD Signals in the $K\pi$ Sector.\label{tab:dcsd}}
\vspace{0.4cm}
\begin{center}
\begin{tabular}{l|c} 
Experiment & $r_{K\pi}^{\mbox{\footnotesize {\rm DCSD}}}$ \\ \hline\hline 
CLEO  & $(0.77 \pm 0.25 \pm 0.25)~\%$ \\
E791  & $(0.68 {}^{\ + 0.34}_{\ -0.33} \pm 0.07)~\%$ \\
\hline\hline
\end{tabular}
\end{center}
\end{table}
\section{Current Status of Exotic Physics Studies}\label{sec:exotic}
Suffice it to say that, despite a number of experimental 
searches, no truly exotic signals have yet been discovered 
in the charm sector.  In the following, we shall touch on 
both the experimental and theoretical situations and 
indicate that at least some theoretical predictions are not 
terribly far from existing limits.  
\subsection{Experimental Studies}\label{subsec:dcsd}
The only signal observed in this category is the detection 
of doubly-Cabibbo-suppressed decay (DCSD), 
and one is clearly stretching things by calling this 
unambiguous SM phenomenon `exotic'.  A number of searches 
for FCNC processes have been carried out, but with null results.  
We shall list a few examples to indicate the present level of 
sensitivity. 
\subsubsection{Observation of Doubly-Cabibbo-Suppressed 
Decay}\label{subsubsec:dcsd}
Although not truly exotic, DCSD occurs with a sufficiently small 
branching ratio to qualify as an impressive experimental 
achievement.  We recall that the context for making the 
measurement is one where DCSD is a background to the more 
interesting possibility of $D^0-{\bar D}^0$ mixing.  Results of 
mixing studies are usually cast in terms of the $r$-parameter, where 
\begin{eqnarray*}
& & r_f^{\mbox{\footnotesize{\rm mix}}} 
\equiv {B_{D^0 \to {\bar D}^0 \to {\bar f}} \over B_{D^0 \to f}} \ ,
\quad \qquad 
r_f^{\mbox{\footnotesize{\rm DCSD}}} \equiv 
{B^{\mbox{\footnotesize{\rm DCSD}}}_{D^0 \to {\bar f}} \over 
B_{D^0 \to f}} \ \ .
\\
& & 
\end{eqnarray*}
In both CLEO~\cite{Ci94} and E791~\cite{E791} experiments, the 
observed wrong-sign 
$K\pi$ signals were interpreted as DCSD events and 
the values displayed in Table~\ref{tab:dcsd} are seen to be 
consistent within error bars.  

DCSD has also been observed in the $K2\pi$ sector 
$D^+ \to K^+ \pi^- \pi^+$ mode by E687~\cite{E687} and by 
E791~\cite{E791a}.  In each case, a branching fraction 
$B_{D^+ \to K^+ \pi^- \pi^+} \simeq 3 \times \tan^4\theta$ 
is observed.  A Dalitz plot analysis of the E791 data 
reveals roughly equal amounts of $K^+\rho^0$, $K^{*0}\pi^+$ 
and nonresonant behavior.

\subsubsection{Searches for FCNC Decays}\label{subsubsec:exsrch}
A survey of searches for exotic decays appears 
elsewhere in these proceedings.  We simply take note 
in Table~\ref{tab:fcnc} that the overall level of sensitivity 
has in some cases reached the ${\cal O}(10^{-6})$ level.    
\begin{table}[t]
\caption{Some Limits on FCNC Decays.\label{tab:fcnc}}
\vspace{0.4cm}
\begin{center}
\begin{tabular}{c|c|c}Mode & Branching Ratio & Experiment \\ \hline\hline
$D^+ \to \pi^+ \mu^+\mu^-$ & $ < 1.8\times 10^{-5}$ &
E791~\cite{Ai96a} \\
$D^0 \to \mu^+ \mu^-$  & $ < 4.2\times 10^{-6}$ & 
E771~\cite{Al96} \\
$D^{0,+} \to X^{0,+}\ell^+ \ell^-$  & 
$ < {\cal O}(10^{-4} \to 10^{-5})$ & CLEO~\cite{Fr96} \\
\hline\hline
\end{tabular}
\end{center}
\end{table}

\subsection{Theoretical Studies of FCNC Processes}\label{subsec:thy2}
Let us suppose that the studies of $D \to M \gamma$ 
and/or $D \to M' \ell^+ \ell^-$ recommended earlier in 
this talk are actually carried out and that some signal is observed.  
Several interpretations are possible:
\begin{itemize}
\item A signal of `new physics',~\cite{sp97} 
\item A SM short-distance effect, akin to the EW-penguin of 
$B \to K^* \gamma$, 
\item A SM long-distance effect.~\cite{bghp97}    
\end{itemize}
In the following, we explore this important issue with reference to the 
$D \to M \gamma$ and $D \to M' \ell^+ \ell^-$ transitions.  
\subsubsection{Weak Radiative Decay ($D \to M \gamma$)}\label{subsec:radweak}
The current experimental limit is $B_{D^0 \to \rho^0,\omega^0 \gamma} 
= {\cal O}(10^{-4})$.  Let us compare this with theoretical
expectations.  The charm analog of the short-distance 
EW penguin vertex $b\to s \gamma$ is $c\to u \gamma$.  
Omitting QCD corrections, one finds a tiny branching ratio 
$B_{c \to u\gamma}^{\rm No~QCD} \simeq 1.4 \times 10^{-17}$.  
Although a study of one-loop QCD corrections~\cite{BuGoHePa96}  
finds a large enhancement, the branching ratio 
$B_{c \to u\gamma}^{\rm 1-loop} \simeq 5 \times 10^{-12}$   
is still tiny.  The same is true even when two-loop QCD corrections 
are included~\cite{Gre96}.  However, two studies of {\it long 
distance effects} (using respectively vector-dominance 
arguments~\cite{BuGoHePa96} and a weak-annihilation/QCD-sum-rule 
approach~\cite{Kh95}) obtain branching ratios at the 
${\cal O}(10^{-6})$ level.  One must therefore keep in mind that 
misinterpretation of observed FCNC signals is possible unless 
long range effects are first understood!

\subsubsection{Weak Decay into Lepton Pairs 
($D \to M' \ell^+ \ell^-$)}\label{subsub:rare}
The process $D \to \pi \ell^+ \ell^-$ is 
analogous to the weak radiative decay just 
considered except that the photon is now
virtual (we denote a virtual 
photon as $\gamma^*$) and must carry sufficient momentum 
transfer to produce an on-shell lepton pair.  
The branching ratio from the short-distance (EW penguin) 
vertex $c \to u \gamma^*$ in the literature is 
${\cal B}_{D\to \pi \ell^+ \ell^-}^{\rm EW~pen}
= {\cal O}(10^{-8})$.  However, there is an almost 
trivial source of lepton pairs arising from the decay 
chain 
\begin{equation}
D \to \pi + \phi \to \pi + \ell^+ \ell^- \ \ .
\end{equation}
Using as input the measured branching ratio $B_{D^+ \to \pi^+ \phi^0} 
= (6.1 \pm 0.6)\times 10^{-3}$, it is estimated from integrating 
$d\Gamma_{D^+ \to \pi^+ \ell^+ \ell^-}/dm^2_{\ell^+\ell^-}$ 
over the $\phi$ peak that a `resonance' branching ratio 
$B^{\rm res}_{D^+ \to \pi^+ \phi^0} \simeq 0.8 \times 10^{-6}$ 
occurs and that the wings of the Breit-Wigner profile of the 
$\phi$ lead to branching ratios ${\cal O}(10^{-7})$, the 
precise magnitudes of which depend on cuts in 
$d\Gamma_{D^+ \to \pi^+ \ell^+ \ell^-}/dm^2_{\ell^+\ell^-}$.~\cite{Si97}  

\section{CP Violation in the $D$ System}
Although there is not time to fully review this 
subject, I consider two matters of special interest and 
worthy of comment.
\subsection{Time Dependence in ${\overline D}^0 - D^0$ 
Mixing}\label{sec:time}
An approximate but accurate formula for 
the time-dependent decay rate for the transition of $D^0$ 
to some final state $f$ is~\cite{Li95,Bl95} 
\begin{equation}
\Gamma_{D^0 (t) \to f} \propto e^{-\Gamma_D t}~\bigg[ X + Y~t + Z~t^2
\bigg]\ \ ,
\label{ti}
\end{equation}
where $x_D, y_D \ll 1$ and $|\lambda| \ll 1$ ($\lambda$ is defined below). 
The constant term $X$ within the brackets arises from 
DCSD, whereas the quadratic term $Z$ is entirely due
to mixing.  The linear term $Y$ arises from interference and 
can be written as 
\begin{equation}
Y = 2 {\cal R}e~\lambda ~\Delta \Gamma_D  + 
   4 {\cal I}m~\lambda ~\Delta m_D  \ \ ,
\end{equation}
where $\lambda$ is a complex number defined by 
$\lambda \equiv {p \over q}~{A \over B}$ with $p$ and $q$ being 
the usual mass matrix parameters and 
\begin{equation}
A \equiv \langle f | H_{\rm wk} | D^0 \rangle \ , \qquad 
B \equiv \langle f | H_{\rm wk} | {\bar D}^0 \rangle \ .
\end{equation}
If one neglects the presumably small $\Delta\Gamma_D$ contribution to 
the $Y$ term, the remaining contribution 
is proportional to ${\cal I}m~\lambda$, which 
can be nonzero if (i) CPV is present (thus inducing a phase in $p/q$) 
and/or (ii) the FSI are different in $D^0 \to f$ and ${\bar D}^0 
\to f$ (thus inducing a phase in $A/B$).  Blaylock {\it et al} 
performed a careful analysis of the time dependence in mixing 
and discussed the importance of allowing for FSI effects in any 
model-independent experimental analysis.~\cite{Bl95}  
Table~\ref{tab:nlmix} displays the mixing values 
obtained from nonleptonic decays ($D \to K\pi$ and $D\to K 3\pi$) 
and obtained under the most general of conditions, 
{\it i.e.} making no assumptions regarding the 
absence of CPV.  The $90\%$~C.L. upper bounds are 
$r_{K\pi,K3\pi}^{\rm mix}({\bar D}^0 \to D^0) < 0.74\%$ and 
$r_{K\pi,K3\pi}^{\rm mix}(D^0 \to {\bar D}^0) < 1.45\%$.   
It is striking that despite the very large data sample of 
E791, the resulting bound on $D^0 - {\bar D}^0$ mixing 
is of the same order-of-magnitude as previously determined.  
Part of the reason for this is the careful attention paid 
by E791 to the possibility of an interference term.
\begin{table}[t]
\caption{Mixing Signals.\label{tab:nlmix}}
\vspace{0.4cm}
\begin{center}
\begin{tabular}{l|c} 
Mixing & $r_{K\pi,K3\pi}^{\rm mix}$     \\ \hline\hline
${\bar D}^0 \to D^0$ & $(0.18^{\ +0.43}_{\ -0.39} \pm 0.17)\%$ \\ 
$D^0 \to {\bar D}^0$ & $(0.70^{\ +0.58}_{\ -0.53} \pm 0.18) \%$ \\ 
\hline\hline 
\end{tabular}
\end{center}
\end{table}

\subsection{CP Violating Asymmetries in $D^\pm$ Decay}\label{sec:asymm}
In meson systems, one can seek CPV effects either via mixing
(indirect CPV) or not (direct CPV).  For $D$ mesons, 
it would appear that the former can be observed only 
from some kind of new physics mechanism.  For the purpose of testing 
CPV in the Standard Model, the search for asymmetries in $D^\pm$ 
decays is particularly important.  The E791 results given in 
Table~\ref{tab:acpv} reveal that the current sensitivity of 
such searches is in the ${\cal O}(10^{-2})$ range.
\begin{table}[t]
\caption{E791 Limits on CPV Asymmetries.\label{tab:acpv}}
\vspace{0.4cm}
\begin{center}
\begin{tabular}{c|c|c}Mode & $a_{\rm CP}$ & $90\%$ CL Limits ($\%$) \\ \hline\hline
$K^-K^+\pi^+$ & $-0.014 \pm 0.029$ & $-6.2 < a_{\rm CP} < 3.4$ \\ 
$\phi\pi^+$ & $-0.028 \pm 0.036$ & $-8.7 < a_{\rm CP} < 3.1$ \\ 
${\bar K}^{*0}(892)K^+$ & $-0.010 \pm 0.050$ & $-9.2 < a_{\rm CP} < 7.2$ \\ 
$\pi^-\pi^+\pi^+$ & $-0.017 \pm 0.042$ & $-8.6 < a_{\rm CP} < 5.2$ \\ 
\hline\hline
\end{tabular}
\end{center}
\end{table}

As regards theoretical predictions, it has been suggested that 
the $\rho\pi$ mode would be particularly attractive,~\cite{Bu95}  
\begin{equation}
a_{\rm CP}^{(\rho\pi)} \equiv {\Gamma_{D^+\to \rho^0\pi^+} 
- \Gamma_{D^- \to \rho^0\pi^-} \over 
\Gamma_{D^+\to \rho^0\pi^+} + \Gamma_{D^- \to \rho^0\pi^-}} 
\simeq -2 \times 10^{-3} \ \ .
\end{equation}
More generally, this reference predicts a number of CPV asymmetries at the 
${\cal O}(10^{-3})$ level.  This is curiously near the 
maximum expected from the SM as estimated out by Burdman,~\cite{gb95} 
\begin{eqnarray}
|a_{\rm CP}^{\rm (charm)}| &\sim& {{\cal I}m~\left[ 
V_{\rm cd} V^*_{\rm ud} V_{\rm cs} V^*_{\rm us} \right] 
\over \lambda^2} {P \over S} \sin \delta_{\rm str} \nonumber \\
&\sim& A^2 \eta \lambda^4 {P \over S} \sin \delta_{\rm str} 
\le 10^{-3} \ \ ,
\end{eqnarray}
where $\lambda$, $A$ and $\eta$ are the Wolfenstein parameters 
of the CKM matrix.  

\section{Theory of ${\overline D}^0 - D^0$ Mixing in the 
Standard Model}\label{sec:Dmix}
In the following, we give an up-to-date and reasonably thorough 
review of attempts to theoretically determine $|\Delta M_D|$ and 
$|\Delta \Gamma_D|$.  
\subsection{Short Distance (`Box') Contribution}\label{subsec:box}
Analytic formulae for the box contributions to $|\Delta M_D|$ 
and $|\Delta \Gamma_D|$ first appeared in the literature over a decade 
ago,~\cite{hag} and were given in approximate form for 
$D^0 - {\bar D}^0$ mixing shortly thereafter.~\cite{Da85}  
The precise formulae are 
\bea
\Delta M_D^{\rm box} &=& {2 \over 3\pi^2} X_D
{(m_s^2 - m_d^2)^2 \over m_c^2} \left[ 
1 - {5\over 4} {B_D' \over B_D} {M_D^2\over (m_c + m_u)^2} \right] 
\nonumber \\
\Delta \Gamma_D^{\rm box} &=& {4 \over 3\pi} X_D 
{(m_s^2 - m_d^2)^2 \over m_c^2} {m_s^2 + m_d^2 \over m_c^2} 
\left[ 1 - {5\over 2} {B_D' \over B_D} {M_D^2\over (m_c + m_u)^2}
\right] \ ,
\eea
where the $b$-quark contribution has been neglected, 
$X_D$ is given by $X_D \equiv \xi_d^2 B_D G_F^2 M_D F_D^2$, 
with $\xi_d \equiv V^*_{cd}V_{ud}$ and no QCD corrections 
are yet included.  The squared GIM suppression factors 
$(m_s^2 - m_d^2)^2 /m_c^2$ indicate the presence of two 
$|\Delta C| = 1$ transitions and there is an additional suppression 
factor $(m_s^2 + m_d^2) / m_c^2$ for $\Delta \Gamma_D$.   
Until recently, the `B-parameters' $B_D$ and $B_D'$ represented a major
source of numerical uncertainty, even as to the sign of the effect.  
However, a very recent lattice determination~\cite{gbs} of both the 
B-parameters has alleviated this source of numerical
ambiguity, leaving the value of $m_c$ as perhaps the least 
well-determined quantity.  We display the $m_c$-dependence of 
the box contributions to $\Delta M_D$, $\Delta \Gamma_D$ 
and $r_{\rm box} \equiv \Delta \Gamma_D / \Delta M_D$ 
in Table~\ref{tab:box}. 
\begin{table}[t]
\caption{Box Contributions to $\Delta M_D$ and $\Delta 
\Gamma_D$.\label{tab:box}}
\vspace{0.4cm}
\begin{center}
\begin{tabular}{c|c|c|c} 
$m_c$~(GeV) & $\Delta M_D$~(GeV) & $\Delta \Gamma_D$~(GeV) 
& $r_{\rm box}$ \\ \hline\hline
$1.30$ & $-0.191 \times 10^{-16}$ & $-0.752 \times 10^{-17}$ & $0.395$ \\
$1.32$ & $-0.176 \times 10^{-16}$ & $-0.681 \times 10^{-17}$ & $0.388$ \\
$1.34$ & $-0.162 \times 10^{-16}$ & $-0.618 \times 10^{-17}$ & $0.381$ \\ 
$1.36$ & $-0.150 \times 10^{-16}$ & $-0.561 \times 10^{-17}$ & $0.375$ \\
$1.38$ & $-0.138 \times 10^{-16}$ & $-0.510 \times 10^{-17}$ & $0.369$ \\
\hline\hline
\end{tabular}
\end{center}
\end{table}

\subsection{Heavy Quark Effective Theory Analysis}\label{subsec:hqet}
The effect of QCD radiative corrections on the box 
contribution has been studied in the context of the 
heavy quark effective theory (HQET).  Georgi gave the original 
formulation of HQET to $D^0 - {\bar D}^0$ mixing,~\cite{Ge92} and 
analysis of the QCD effects appeared soon thereafter.~\cite{Oh93} 
HQET represents the most natural means for computing QCD 
radiative corrections for heavy quark processes, as otherwise 
uncontrollable logarithmic factors are avoided.  There are several 
steps in this approach: (i) obtain an effective hamiltonian 
in terms of local operators (OPE), (ii) run the energy scale 
(RG equations) from $\mu \simeq M_{\rm W}$ down to the value of 
interest ($\mu \simeq m_c$ in our case) and (iii) obtain 
accurate values for the matrix elements of the local operators.  
For application to $D^0 - {\bar D}^0$ mixing, 
operators of dimension four, six and eight were considered, 
and the maximum effect obtained was estimated to be $|\Delta M_D| \le 3.5 
\times 10^{-17}~{\rm GeV}$.~\cite{Oh93}   The main uncertainites here 
involve convergence of the OPE ({\it i.e.} is the $c$-quark 
really heavy?) and evaluation of the multi-quark matrix 
elements (effects of nonperturbative physics).  

\subsection{Dipenguin Contribution}\label{subsec:dipen}
The box contribution is not the only short distance 
contribution.  It has long been understood that the 
effect of two penguin operators (dipenguin) can also occur as 
a four-quark local operator.~\cite{dgv}  This effect 
has been studied for both the $K^0 - {\bar K}^0$ and 
$B^0 - {\bar B}^0$ mixings, but until recently not 
for $D^0 - {\bar D}^0$ mixing.  This gap in the 
literature has been filled by a very recent calculation 
of Petrov who finds~\cite{Pe97}  
\begin{equation}
|\Delta M_D^{\rm di-pen}| \le 10^{-17}~{\rm GeV} \ \ .
\end{equation}
Thus, the dipenguin does not lead to a surprisingly 
large contribution.  

\subsection{Long Distance (`Dispersive') Contributions}\label{subsec:ld}
It could happen at the relatively light charm mass scale that 
significant corrections to the short-distance box contribution 
exist.  It is difficult to obtain an accurate determination 
of this long distance component, but nonetheless instructive 
to study the various categories.

\subsubsection{Single-particle Sector}\label{subsec:single}
Although the analysis of the single particle sector was carried 
out some time ago,~\cite{bghp94} this represents its first 
published appearance.  Individual contributions are displayed in 
Table~\ref{tab:Dmix} for a specific choice $\theta_P = - 19^o$ 
of the $\eta - \eta'$ mixing angle, and the 
dependence on $\theta_P$ is given in the two rightmost columns.  
\begin{table}[t]
\caption{One-particle Contributions to $\Delta m_D$.\label{tab:Dmix}}
\vspace{0.4cm}
\begin{center}
\begin{tabular}{c|c||c|c} 
Mode & $\Delta m_D~({\rm GeV})$ & 
$\theta_P$~({\rm deg}) & $\Delta m_D$~({\rm GeV}) \\ \hline\hline
$K$ & $0.1345\cdot 10^{-15}$ & 
$-13$ & $0.4127\cdot 10^{-16}$ \\
$B_d$ & $-0.1830\cdot 10^{-17}$ &
$-14$ & $0.3948\cdot 10^{-16}$ \\
$B_s$ & $0.2407\cdot 10^{-17}$ &
$-15$ & $0.3757\cdot 10^{-16}$ \\
$\pi$ & $-0.1545\cdot 10^{-17}$ &
$-16$ & $0.3554\cdot 10^{-16}$ \\
$\eta_B$ & $0.7138\cdot 10^{-21}$ & 
$-17$ & $0.3339\cdot 10^{-16}$ \\
$\eta$ & $-0.7227\cdot 10^{-16}$ &
$-18$ & $0.3112\cdot 10^{-16}$ \\
$\eta'$ & $-0.3250\cdot 10^{-16}$ &
$-19$ & $0.2875\cdot 10^{-16}$ \\
Total & $0.2875\cdot 10^{-16}$ &
$-20$ & $0.2627\cdot 10^{-16}$ \\ \hline\hline
\end{tabular}
\end{center}
\end{table}

\subsubsection{Multi-particle Sector}\label{subsec:multi}
Historically, the first of the long distance components 
to be studied was the two-particle charged-pseudoscalar 
sector.~\cite{Wo85,DoGoHo86}  The motivation was then 
(and remains to this day) quite clear -- the large $SU(3)$ 
breaking evidenced in $D$ decays.  One example of this 
concerns the ratio $\Gamma_{D^0 \to K^+K^-}/\Gamma_{D^0 \to 
\pi^+ \pi^-} \simeq 3$, which is unity in the $SU(3)$ limit.  
Chau and Chang have analyzed several related decays and 
find the large breaking to be an accumulation of a number 
relatively minor effects whose ultimate impact is 
large.~\cite{cc94}  

In the original quantitative study of the long distance two-particle 
sector,~\cite{DoGoHo86} an approximate 
correspondence $|\Delta M_D^{\rm disp}/\Delta \Gamma_D|\sim 0.2$ 
appeared natural.  If we combine this with a 
recent analysis~\cite{Bu95} which estimates 
$\Delta \Gamma_D /\Gamma_D \simeq 10^{-3}$ 
via an explicit sum over two-particle $PP, VP, SP$ 
modes, we obtain the following order-of-magnitude result,
\begin{eqnarray}
|\Delta M_D^{\rm disp}| &\sim& 0.2 ~\Delta \Gamma_D \qquad (\rm heuristic \ 
assumption) \nonumber \\
&=& 0.2 ~{\Delta \Gamma_D \over \Gamma_D}~ \Gamma_D 
\nonumber \\
&\sim& 0.2 \times 10^{-3} \times 10^{-12}~{\rm GeV} 
\sim {\cal O}(10^{-16}~{\rm GeV}) \ \ .
\label{multi}
\end{eqnarray}
Although this order-of-magnitude estimate is far from rigorous, 
it represents a reasonable bound on the 
magnitude of the dispersive component.   

\section{New Physics}
Although a variety of important SM tests remain to be carried out, 
increasing attention is being given to probes of 
new physics.  In the following, we recall 
some recent literature regarding the time-dependent decay rate 
$\Gamma_{D^0 (t) \to f}$ and then catalog a number of potential 
new physics contributions to $\Delta M_D$.  
\subsection{Time Dependence in ${\overline D}^0 - D^0$ 
Mixing}\label{sec:newtime}
Several papers have pointed out the potential 
importance of the interference term ({\it cf}~the $Y$-term 
in Eq.~(\ref{ti})) for studies of mixing in the 
$D$ sector.  First, Wolfenstein noted that detection of mixing 
at current levels of sensitivity would indicate new physics, 
and that in the absence of FSI, the effect would 
involve CPV.~\cite{Wo95}  Papers by Blaylock, Seiden $\&$ 
Nir~\cite{Bl95} and by Browder $\&$ Pakvasa~\cite{Br96}  
provided a more detailed look at new physics possibilities.  
The latter also stressed that the term $Y$ would survive (i) for zero CPV 
in the combination $\Gamma_{D^0 (t) + {\bar D}^0 (t)}$ and 
(ii) for zero FSI in the combination $\Gamma_{D^0 (t) - 
{\bar D}^0 (t)}$.  A summary of CPV contributions to 
$D^0 - {\bar D}^0$ mixing from a variety of new physics scenarios 
exists in preprint version.~\cite{BuGoHePa95}. 

\subsection{Possible Contributions to ${\overline D}^0 - D^0$ 
Mixing}\label{sec:new}
\begin{figure}
\vskip .1cm
\hskip 1cm
\psfig{figure=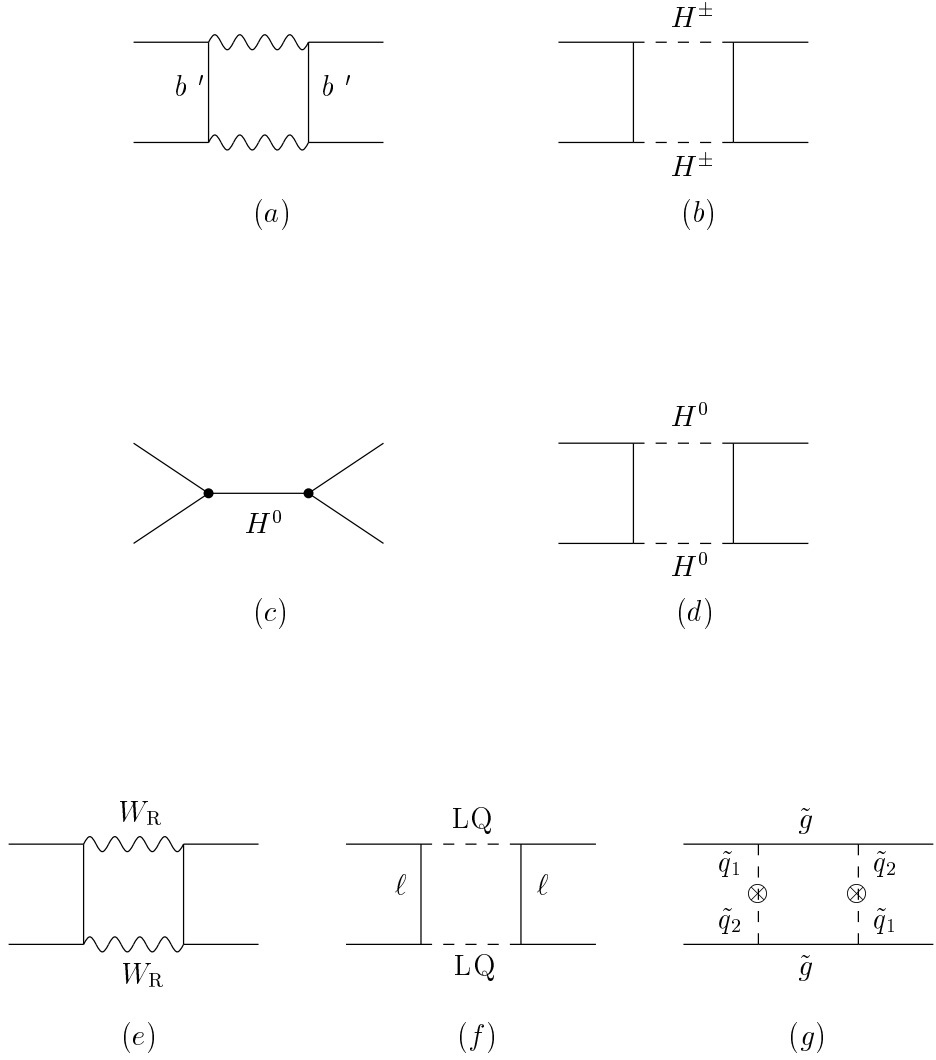,height=5.5in}
\caption{Effects of new physics on ${\bar D}^0D^0$ Mixing:
(a) extra $Q = -1/3$ quark $b~'$, (b) charged higgs scalars $H^\pm$, 
(c-d) tree, box contributions of flavor-changing neutral higgs 
scalars $H^0$, (e) left-right symmetric $W$-boson $W_R$, (f) 
leptoquark LQ, (g) gluino ${\tilde g}$ and 
squarks ${\tilde q}_{1,2}$.\hfill 
\label{fig:models}}
\end{figure}

{\it A propos} of new physics contributions to a given system, 
there is by now a well-developed path to follow:
\begin{itemize}
  \item Pick some Model-X of new physics, identifying its degrees 
of freedom and the extent of its parameter space, 
  \item Calculate its effect on a given observable, 
  \item Perform a phenomenological analysis and decide whether 
  \subitem (a) Model-X contribution unobservable?
  \subitem (b) Model-X parameter space constrained?
  \subitem (c) Model-X ruled out?
\end{itemize}
Table~\ref{tab:newposs} displays a rich spectrum of new 
physics possibilities for contributing to 
$D^0 - {\bar D}^0$ mixing.  Indeed, all the models typically 
discussed in the literature are available.  We 
display some of the mechanisms for inducing $D^0 - {\bar D}^0$ 
mixing in Figure~\ref{fig:models}.  

The remainder of the procedure -- to calculate the effects 
over the available parameter space -- is reviewed ably by 
Hewett~\cite{he95} and by Burdman,~\cite{gb95} who also 
refer to the original literature.  We simply note here that:

\ \ (i) Although the HERA-motivated interest in leptoquarks 
is quite recent, there already exist interesting 
phenomenological studies of such particles,
including their effect on $D^0 - {\bar D}^0$.~\cite{dbc93}

\ (ii) In the flavor-changing neutral 
Higgs scenario, it has recently been pointed out 
that a reasonably modest improvement in the bound 
on $D^0 - {\bar D}^0$ mixing will make this test 
competitive with mixing constraints in the kaon and 
$B_d$ systems.~\cite{At97} 

(iii) The tight experimental bounds on FCNC 
processes place severe restrictions in supersymmetric models 
on the structure of sfermion mass matrices.  Several approaches 
(`flavor universality', `alignment', `gauge mediation') 
are available for dealing with this problem.  We note 
a recent `mass insertion approximation' analysis of FCNC 
constraints which is relatively model independent and 
includes a study of $D^0 - {\bar D}^0$ mixing.~\cite{Ga97} 

\begin{table}[t]
\caption{New Physics Possibilities.\label{tab:newposs}}
\vspace{0.4cm}
\begin{center}
\begin{tabular}{l|l}New Physics & Comment \\ \hline\hline
Extra quarks & $4^{th}$ family weak-isodoublet ($t', b'$) \\ 
             & singlets ($Q = -1/3\ {\rm or}\ Q = + 2/3$)\\
Extra scalars & charged higgs $H^\pm$ \\ 
              & neutral (FC) higgs ($H^0$) \\ 
Extra W-bosons & left-right symmetry ($W_R$) \\
Leptoquarks &  \\
Sparticles & R-parity conserving \\
            & R-parity violating \\
Compositeness & techniparticles \\
Family symmetry &     \\
\hline\hline
\end{tabular}
\end{center}
\end{table}

\section{Summary}
We have argued for using B-factory capabilities to carry out 
charm physics experiments in the three areas of rare decay searches, CPV 
asymmetry searches and improvement of the bound on $D^0-{\bar D}^0$ 
mixing.  Each of these has greater intrinsic interest 
than more conservative choices.  To support this view, 
an up-to-date summary of these topics has been provided, and several 
new results (particularly in the discussion of $D^0-{\bar D}^0$ 
mixing) have been presented.  Additional studies are 
underway and theoretical advances in the near term are anticipated.  
We end with the reminder that there are advantages to doing 
$D^0-{\bar D}^0$ mixing searches at an asymmetric $B$-factory, 
particularly using a hadronic tag and a semileptonic 
mixing signal.~\cite{Bl97}

\section*{Acknowledgments}
Research by the UMass theory group is supported by 
the National Science Foundation.  The input of my 
collaborators Gustavo Burdman, JoAnne Hewett and 
Sandip Pakvasa and of Mark Windoloski is gratefully acknowledged.

\end{document}